\documentclass{jetpl}
\twocolumn

\usepackage{amsmath,amssymb,epsfig,color,pstricks,bm,graphics,alltt}

\begin{document}

\lat

\title{High Temperature Superconductivity in Transition Metal Oxypnictides: 
         a Rare-Earth Puzzle?}

\rtitle{Rare Earth Puzzle}

\sodtitle{High Temperature Superconductivity in Transition Metal Oxypnictides: 
         a Rare-Earth Puzzle?}

\author{I.\ A.\ Nekrasov$^+$, Z.\ V.\ Pchelkina$^*$, M.\ V.\ Sadovskii$^+$}

\rauthor{I.\ A.\ Nekrasov, Z.\ V.\ Pchelkina, M.\ V.\ Sadovskii}

\sodauthor{Nekrasov, Pchelkina, Sadovskii }

\sodauthor{Nekrasov, Pchelkina, Sadovskii }

\address{$^+$Institute for Electrophysics, Russian Academy of Sciences, 
Ural Division, 620016 Yekaterinburg Russia}

\address{$^*$Institute for Metal Physics, Russian Academy of
Sciences, Ural Division, 620041 YekaterinburgGSP-170, Russia}

\dates{Today}{*}

\abstract{
We have performed an extensive {\it ab initio} LDA and LSDA+U calculations of electronic 
structure of newly discovered high-temperature superconducting series 
ReO$_{1-x}$F$_x$FeAs (Re=La,Ce, Pr, Nd, Sm and hypothetical case of Re=Y). 
In all cases we obtain almost identical electronic spectrum (both energy 
dispersions and the densities of states) in rather wide energy interval 
(about 2 eV) around the Fermi level.
We also debate that this fact is unlikely to be changed by the account of strong correlations.
It leads inevitably to the same critical temperature $T_c$  of superconducting
transition in any theoretical BCS-like mechanism of Cooper pair formation.
We argue that the experimentally observed variations of $T_c$ for different
rare-earth substitutions are either due to disorder effects or less probably because of possible changes
in spin-fluctuation spectrum of FeAs layers caused by magnetic interactions with
rare-earth spins in ReO layers.
}

\maketitle
The recent discovery of the new superconductor LaO$_{1-x}$F$_x$FeAs with the
transition temperature $T_c$ up to 26K \cite{kamihara_08,chen,zhu,mand} 
has been immediately followed by the reports of even higher $T_c=$41K in
CeO$_{1-x}$F$_x$FeAs \cite{chen_3790}, $T_c=$43K in SmO$_{1-x}$F$_x$FeAs 
\cite{chen_3603} and $T_c=$52K in NdO$_{1-x}$F$_x$FeAs and 
PrO$_{1-x}$F$_x$FeAs \cite{ren_4234,ren_4283}. A number of other element
substitutions were studied producing compounds with rather wide interval of
$T_c$ values \cite{chen_4384}. These discoveries of the whole
new class of superconductors, based on doped layered
oxypnictides ReOMPn (Re=La,Ce,Pr,Nd,Sm,Gd; M=Mn,Fe,Co,Ni,; Pn=P,As), open the new
chapter in studies of high-temperature superconductivity outside the well-known 
domain of copper oxides. 
Of course, at the moment, the microscopic nature of superconductivity in these
compunds remains unclear, though a number of aspects of this has already been
discussed \cite{dolg,mazin,aoki,dai,han,eschr}.

The LDA electronic structure and phonon spectrum of LaFeAsO were first
calculated in Ref. \cite{singh} (see also \cite{dolg}). These were extended
to an account of strong electronic correlations on Fe via LDA+DMFT in Ref. 
\cite{haule}. Spin polarized LDA calculations for LaFeAsO with 
antiferromagnetic ordering in the ground state were done in 
Refs. \cite{hirsch,ma}. First principles (FLAPW) calculations of the doping
dependent phase diagram of LaOMAs (M=V-Cu) were performed in Ref. \cite{xu}.
 
Here we present the results of an extensive {\it ab initio} calculations of 
electronic structure of newly discovered high-temperature superconducting 
series ReO$_{1-x}$F$_x$FeAs (Re=La, Ce, Nd, Pr, Sm and hypothetical case of 
Re=Y) within LDA and LSDA+U frameworks. 


\begin{table}[htb]
\footnotesize
\caption{Table 1. Combined list of experimental and calculated
parameters for ReOFeAs systems.}
\label{tab1}
\begin{tabular}{|l|c|c|c|c|c|c|}
\hline
ReOFeAs   &La&Ce&Pr&Nd\\
\hline
T$_c$, K  &26    &41    &52    &51.9      \\
$a$, \AA  &4.035334&3.996&3.9853&3.940\\
$c$, \AA &8.740909&8.648&8.595 &8.496\\
Source     &Ref.~\cite{kamihara_08}&Ref.~\cite{chen_3790}&Ref.~\cite{ren_4283}&Ref.~\cite{chen_3603}\\
LDA N$_{E_f}$,&56.19&56.51&52.1&54.21\\
states/Ry/cell&&&&\\
\hline
\end{tabular}
\end{table}
\normalsize

\begin{figure}[htb]
\includegraphics[clip=true,width=0.45\textwidth]{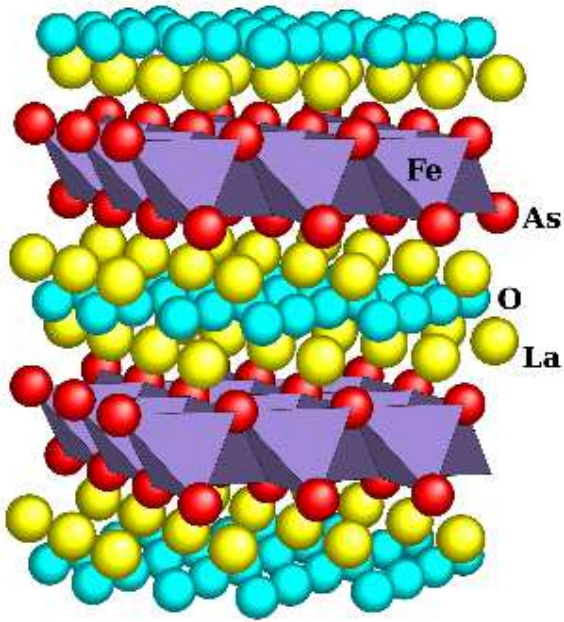}
\label{struct}
\caption{Fig. 1. Tetragonal crystal structure of LaOFeAs compound. 
FeAs layers are formed by Fe ions tetrahedraly surrounded by As ions.}
\end{figure}
  According to the experimental data all ReOFeAs (Re=La, Ce, Pr, Nd, Sm) compounds have (at room temperature)
  well defined tetragonal ZrCuSiAs-type crystal structure with the P4/nmm space group. 
  The structure of iron oxypnictides is formed by FeAs layers sandwitched between LaO layers. 
  These two layers are separated by a distance of 1.8 \AA. The nearest neighbors of Fe ions are four 
  As ions which form a tetrahedron. The crystal structure of LaOFeAs is shown in Fig.~1. 
  The values of experimentally obtained lattice constants for different known rare-earth substitutions 
  are presented in Table~1.
  As the atomic number of rare-earth element increases the lattice
  constants decrease indicating the contraction of the cell due to the lanthanoid compression effect. 
  The systematic decrease of the lattice constants was also observed for F$^-$ substitution to the 
  site of O$^{2-}$ ions \cite{kamihara_08}. 
  
  The electronic structure of ReOFeAs (Re=La, Ce, Pr, Nd, Sm and Y) compounds was calculated in the local (spin)
  density approximation (L(S)DA) and LSDA+U~\cite{LSDAU} method by using linearized muffin-tin orbitals (LMTO)
  \cite{LMTO}.
  The details of the atomic spheres radii and LMTO basis set used in the calculation are given in the
  Table~2. Since the experimental data for atomic coordinates are limited, in the case of Ce, 
  Pr, Nd, Sm and Y systems we used the same z$_{Re}$ and z$_{As}$ as for the LaOFeAs compound: 
  z$_{La}$=0.141545, z$_{As}$=0.65122~\cite{kamihara_08}.To our knowledge there is no experimental 
  data for the lattice parameters and atomic coordinates for SmOFeAs.  Thus
  for SmOFeAs we use the same lattice parameters and atomic positions as for Nd system.
  The hypothetical compound with yttrium is not synthesized yet and for its electronic structure calculation we exploit 
  the data for La compound from Ref.~\cite{kamihara_08}. We also keep fixed the radii of atomic spheres of Fe, As and O 
  across the series. The $f$-states of rare-earths were treated as a pseudocore states. The Brillouin zone was sampled 
  using 20 irreducible {\bf k} points. 

\begin{table}[ht]
\caption{Table 2. Details of LMTO calculations for ReOFeAs (Re=La, Ce, Pr, Nd, Sm, Y).}
\label{tab2}
\begin{center}
\begin{tabular}{|l|c|c|}
\hline
 type of atom & LMTO basis     & R$_{AS}$, \AA \\
\hline
La          & 6$s$5$d$4$f$     &1.63	      \\
Ce          & 6$s$5$d$4$f$     &1.61		  \\
Sm          & 6$s$5$d$4$f$     &1.59		  \\
Nd          & 6$s$5$d$4$f$     &1.62		  \\
Pr          & 6$s$5$d$4$f$     &1.625		  \\
Y           & 5$s$4$d$         &1.637		  \\
Fe           & 4$s$4$p$3$d$    &1.377  	   \\
As           & 4$s$4$p$        &1.42  	   \\
O            & 3$s$2$p$        &1.09       \\
\hline
\end{tabular}
\end{center}
\end{table}

\begin{figure}[!h]
\includegraphics[clip=true,width=0.32\textwidth,angle=270]{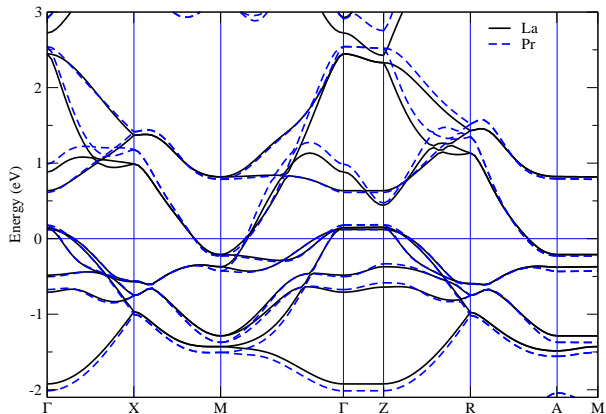}
\label{la_bands}
\caption{Fig. 2. LDA energy bands for LaOFeAs (solid lines) and PrOFeAs (dashed lines) compounds.
The Fermi level corresponds to zero.}
\end{figure}
\begin{figure}[!h]
\includegraphics[clip=true,width=0.5\textwidth,angle=270]{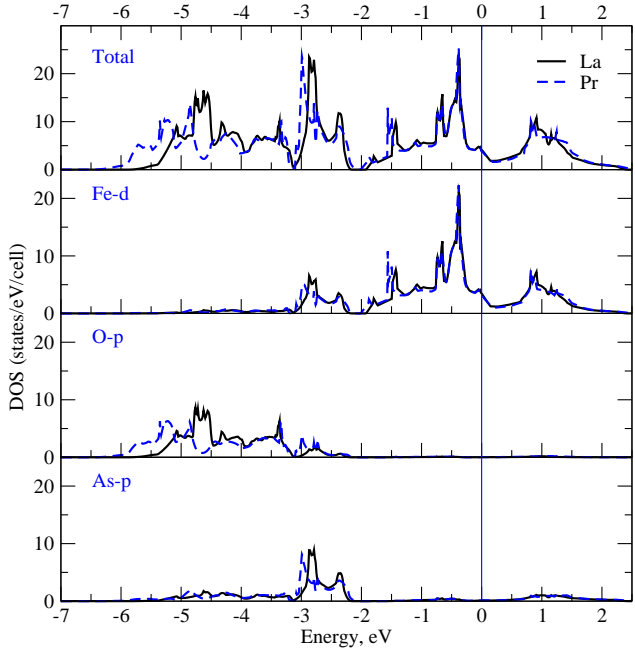}
\label{la_dos}
\caption{Fig. 3. Total and partial LDA DOS for LaOFeAs (solid lines) and PrOFeAs (dashed lines) compounds.
The Fermi level corresponds to zero.}
\end{figure}

  The obtained LDA bands along high symmetry lines of the Brillouin zone and density of states (DOS) 
  for pure LaOFeAs and PrOFeAs superposed with each other
  are shown in Fig.~2 and Fig.~3, correspondingly. Both density of 
  states and band structure are in good agreement with the more accurate calculation within full potential linearized
  augmented planewave (LAPW) method using WIEN2k package~\cite{singh_0429, skorikov}.
  The only significant change in spectra to a substitution of different rare-earths reduces to the
  increase of the size of tetrahedral splitting because
  of lattice contraction. One can see these effects at about -1.5 eV for Fe-$d$ states,
  -3 eV for As-$p$ (Figs.~3,~4).

\begin{figure}[!h]
\includegraphics[clip=true,width=0.5\textwidth,angle=270]{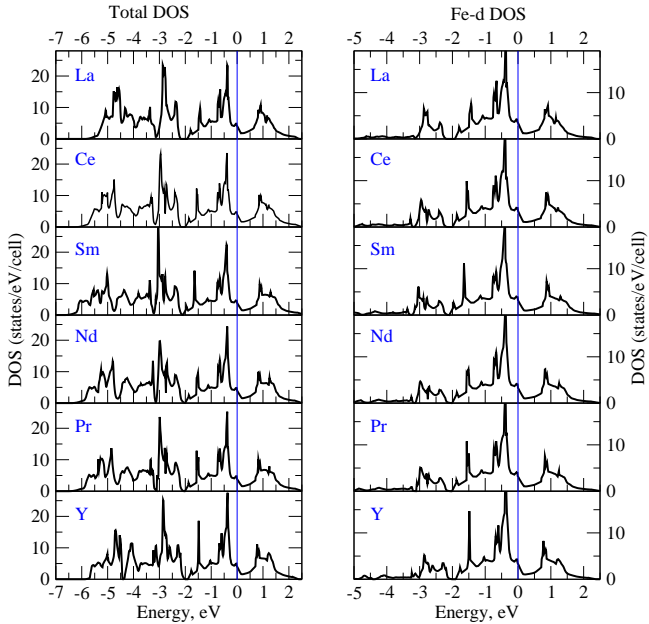}
\label{re_dos}
\caption{Fig. 4. Total (left) and partial Fe-$d$ (right) LDA DOS for series ReOFeAs systems.
The Fermi level corresponds to zero.}
\end{figure}

These results clearly show that electronic structure of ReOFeAs compounds
(e.g. DOS around the Fermi level, see Table~1)
practically does not depend on the kind of rare-earth ion (Re) used in a
wide energy interval around the Fermi level, which is relevant for
superconductivity in FeAs layers. This simply follows from the fact that
electronic states of ReO layers are rather far from the Fermi level. Thus
there is almost negligible hybridization between Fe-$d$ and O-$p$
electronic states as illustrated by partial DOS contributions shown in Fig.~3.
Within FeAs layers however there is quite sizeable hybridization 
between Fe-$d$ and As-$p$. Smaller but still finite hybridization between
O-$p$ and As-$p$ is obtained.
Therefore the ReOFeAs systems can be considered as quasi-two-dimensional
which is similar to the case of conventional high-T$_c$ cuprates. 

Similarity of compounds within ReOFeAs series is unlikely to be changed by the account of strong electronic
correlations on Fe, along the lines of Ref. \cite{haule}, since only bare Fe bands
are important in such type of calculations. In Fig.~4 we show that no 
substantial modifications of Fe-$d$ electronic states close to the Fermi level are found
due to rare-earths substiutions. But one can in principle expect that interplay
of orbital degrees of freedom within Fe-$d$ shell because of different tetrahedral
splitting might affect DOS value on the Fermi level within multi-orbital Hubbard model.
However no such effects has been reported in Ref. \cite{haule}.

\begin{figure}[!h]
\includegraphics[clip=true,width=0.5\textwidth,angle=270]{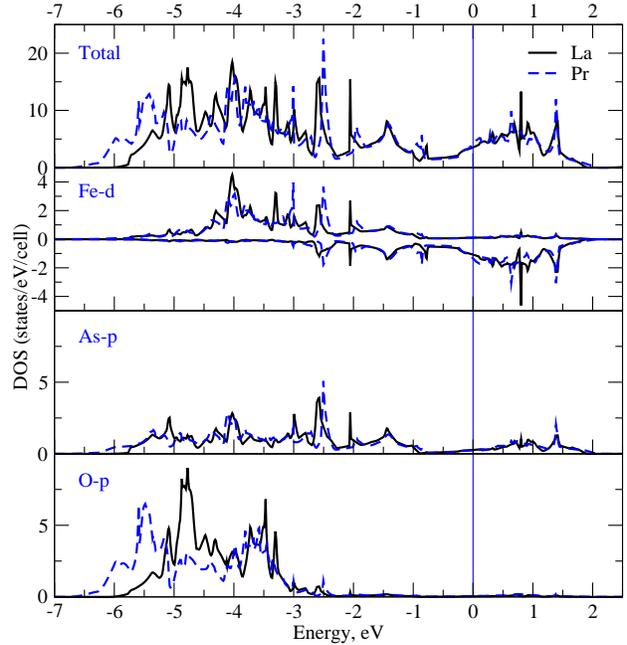}
\label{la_u_dos}
\caption{Fig. 5. Total and partial LSDA+U DOS for LaOFeAs (solid lines) and PrOFeAs (dashed lines) compounds.
The Fermi level corresponds to zero.}
\end{figure}

Also we performed LSDA and LSDA+U calculation for collinear antiferromagnetically ordered Fe ions in FeAs plane.
Spins were directed perpendicular to the FeAs plane. Values of Coulomb interaction $U$
and Hund exchange $J$ were taken 3 and 1 eV respectively.
Corresponding LSDA+U DOS for LaOFeAs and PrOFeAs compounds are presented in Fig.~5.
Here one can say that any kind of spin-polarized
LSDA or LSDA+U type of calculations do not make these compounds remarkably different.
Our results are in good agreement with Refs.~\cite{hirsch,ma}.
Also one can mention that for all systems we obtained half-metallic LSDA+U solution.

We also have done
LSDA+U calculations with AFM ordered Re ions, while Fe ions were not magnetic.
We have found that rather large magnetic moments of Re ions have alsmost no
effect on FeAs layer. Namely no magnetization of FeAs layer induced by
Re ions is observed.


Thus, situation with rare-earth ions in ReOFeAs system seems in many respects
quite similar to that in ReBa$_2$Cu$_3$O$_{7-x}$ series of copper oxide
superconductors, studied in the very early days of high-$T_c$ research
in cuprates \cite{trst,htsc}. In this series, the electronic states of
rare-earth ions also just do not overlap with electronic states of CuO$_2$
conducting planes. This leads to the well established fact of almost complete
independence of superconducting $T_c\sim 92K$ on the type of the rare-earth
ion for the case of Re=Y,Nd,Sm,Eu,Gd,Ho,Er,Tm,Yb,Lu,Dy \cite{trst}, with only
two exceptions -- the much lower value of $T_c\sim 60K$ in the case of La and
no superconductivity in the very special case of Pr \cite{htsc}.

Almost identical electronic structure of Fe oxypnictides 
ReOFeAs with different rare-earth substitutions in a wide energy range around 
Fermi level should inevitably lead to practically the same values of 
superconducting transition temperature T$_c$ in any kind of BCS-like pairing 
mechanism.  Rare-earths just do not influence electronic structure of FeAs 
layers due to a very small overlap of appropriate electronic states. 
Thus, no 
significant change  of pairing interaction constant should be expected. There 
is no serious reason to expect any significant change of phonon's spectrum 
due to the rare-earth substitution either as well as spin fluctuation 
spectrum in FeAs layers.
 
Thus we have a kind of a rare-earth puzzle! In contrast to ReBa$_2$Cu$_3$O$_{7-x}$ 
series of copper oxides experimental data of Refs. \cite{chen_3790,chen_3603,
ren_4234,ren_4283, chen_4384} show that the rare-earth substitutions in 
ReOFeAs series lead to rather wide range of T$_c$ - from
about 10 K in case of Gd up to 52 K in case of Nd and Pr systems. 
Two possible explanations of this puzzle seem feasible at the moment:
 
(i) Different quality of samples (disorder effects) may lead to a wide range 
of T$_c$, as disorder usually leads to a strong suppression of exotic types 
of pairing (e.g. anisotropic $p$- or $d$-wave, triplet pairing, etc.), which
are discussed now for ReOFeAs series \cite{mazin,aoki,dai,han}. In general,
situation here is qualitatively similar to the case of copper-oxides, where
$d$-wave pairing is strongly suppressed by disorder (see e.g. \cite{scloc}). 
Similar argument is usually used to explain typically lower values of $T_c$ 
in LaBa$_2$Cu$_3$O$_{7-x}$ as due to disorder in La and Ba positions and oxygen 
vacancies \cite{htsc}. This stresses the importance of systematic studies of 
disorder effects in the new class of ReOFeAs  high-T$_c$ superconductors.
 
(ii) Basically, spin-ordering effects of rare-earths (like Ce, Pr, Nd, Sm, 
Gd possessing localized spin moment) may, in principle, induce a change of 
spin fluctuation spectrum (and T$_c$) due to magnetic interaction between 
rare-earths and FeAs layers (in case of spin-fluctuation mechanism of 
pairing, as discussed e.g. in Refs. \cite{mazin,aoki}).  However, these 
interactions are most probably rather weak,
as illustrated by our LSDA+U calculations. Still, these effects are worth studying since the 
situation with rare-earth spins ordering is at present unclear from the 
experimental point of view.  

This work is supported by RFBR grants 08-02-00021, 08-02-00712, RAS programs 
``Quantum macrophysics'' and ``Strongly correlated electrons in 
semiconductors, metals, superconductors and magnetic materials'',
Grants of President of Russia MK-2242.2007.2(IN), MK-3227.2008.2(ZP)
and scientific school grant SS-1929.2008.2, interdisciplinary 
UB-SB RAS project, Dynasty Foundation (ZP) and Russian Science Support Foundation(IN).

\end{document}